\documentclass[prb,
twocolumn,
amsmath,
amssymb,
superscriptaddress]{revtex4-2}

\usepackage{graphicx}
\usepackage{bm}
\usepackage{hyperref}
\usepackage{csquotes}
\usepackage[dvipsnames]{xcolor}
\usepackage{siunitx}
\usepackage{tikz}

\providecommand\vect[1]{\bm{#1}}

\begin{document}
\title{Numerical Calculation of the Hopf Index for 3D Magnetic Textures}

\author{R.\ Knapman}
\email{ross.knapman@uni-due.de}
\affiliation{Faculty of Physics and Center for Nanointegration Duisburg-Essen (CENIDE), University of Duisburg-Essen, 47057 Duisburg, Germany}

\author{M.\ Azhar}
\affiliation{Faculty of Physics and Center for Nanointegration Duisburg-Essen (CENIDE), University of Duisburg-Essen, 47057 Duisburg, Germany}

\author{A.\ Pignedoli}
\affiliation{Faculty of Physics and Center for Nanointegration Duisburg-Essen (CENIDE), University of Duisburg-Essen, 47057 Duisburg, Germany}

\author {L.\ Gallard}
\affiliation{Université de Strasbourg, CNRS, Institut de Physique et Chimie des Matériaux de Strasbourg, F-67000 Strasbourg, France}

\author {R.\ Hertel}
\affiliation{Université de Strasbourg, CNRS, Institut de Physique et Chimie des Matériaux de Strasbourg, F-67000 Strasbourg, France}

\author{J.\ Leliaert}
\affiliation{Department of Solid State Sciences, Ghent University, 9000 Ghent, Belgium}

\author{K.\ Everschor-Sitte}
\affiliation{Faculty of Physics and Center for Nanointegration Duisburg-Essen (CENIDE), University of Duisburg-Essen, 47057 Duisburg, Germany}

\begin{abstract}
To gain deeper insight into the complex, stable, and robust configurations of magnetic textures, topological characterisation has proven essential. In particular, while the skyrmion number is a well-established topological invariant for 2D magnetic textures, the Hopf index serves as a key topological descriptor for 3D magnetic structures. In this work, we present and compare various methods for numerically calculating the Hopf index, provide implementations, and offer a detailed analysis of their accuracy and computational efficiency. Additionally, we identify and address common pitfalls and challenges associated with the numerical computation of the Hopf index, offering insights for improving the robustness of these techniques.
\end{abstract}

\maketitle

\section{Introduction}

In the study of topological invariants, the Hopf index plays a crucial role in characterising the topology of 3D vector fields~\cite{Faddeev1997}. In a plethora of physical systems including liquid crystals~\cite{Tkalec2011}, superfluids~\cite{Kleckner2016} and magnetic materials~\cite{Sutcliffe2017}, complex vector field configurations like hopfions arise whose topology is quantified by the Hopf index $H$. 

A tool for numerically computing the Hopf index of a given vector field $\vect{F}$ is the Whitehead integral formula~\cite{Whitehead1947}
\begin{equation}
\label{eq:HopfIndex}
    H =  -\int_V \mathrm{d}^3 r \, \vect{F} \cdot \vect{A},
\end{equation}
where $\vect{A}$ is the gauge potential of the vector field, i.e.\ $\vect{F} = \vect{\nabla} \times \vect{A}$ and $V$ is the volume in which the vector field $\vect{F}$ is defined~\footnote{Please note that $H$ is also often called ``helicity'', but we refrain from introducing this word here to not confuse it with skyrmion helicity which describes the skyrmion in-plane angle.}. Note that, for a vector field $\vect{F}$ to admit a gauge potential $\vect{A}$, the space in which the vector field is defined must be simply connected, and $\vect{F}$ must be solenoidal, i.e.\ $\vect{\nabla}\cdot {\vect F} = 0$.

For 3D magnetic textures, which are described by a normalised vector field $\vect m$, the Hopf index is typically computed for the corresponding emergent magnetic field whose $i$-th component is given by 
\begin{equation}
    \label{eq:EmergentField}
    F_i = \frac{1}{8\pi} \epsilon_{ijk} \vect{m} \cdot \left( \partial_{j} \vect{m} \times \partial_{k} \vect{m}\right),
\end{equation}
where $i,j,k \in \{x,y,z\}$. Note that $\vect \nabla \cdot \vect F =0$ in smooth magnetic textures, and thus a gauge potential exists on a simply connected volume.

\begin{figure*}
    \centering
    \includegraphics[width=0.95\linewidth]{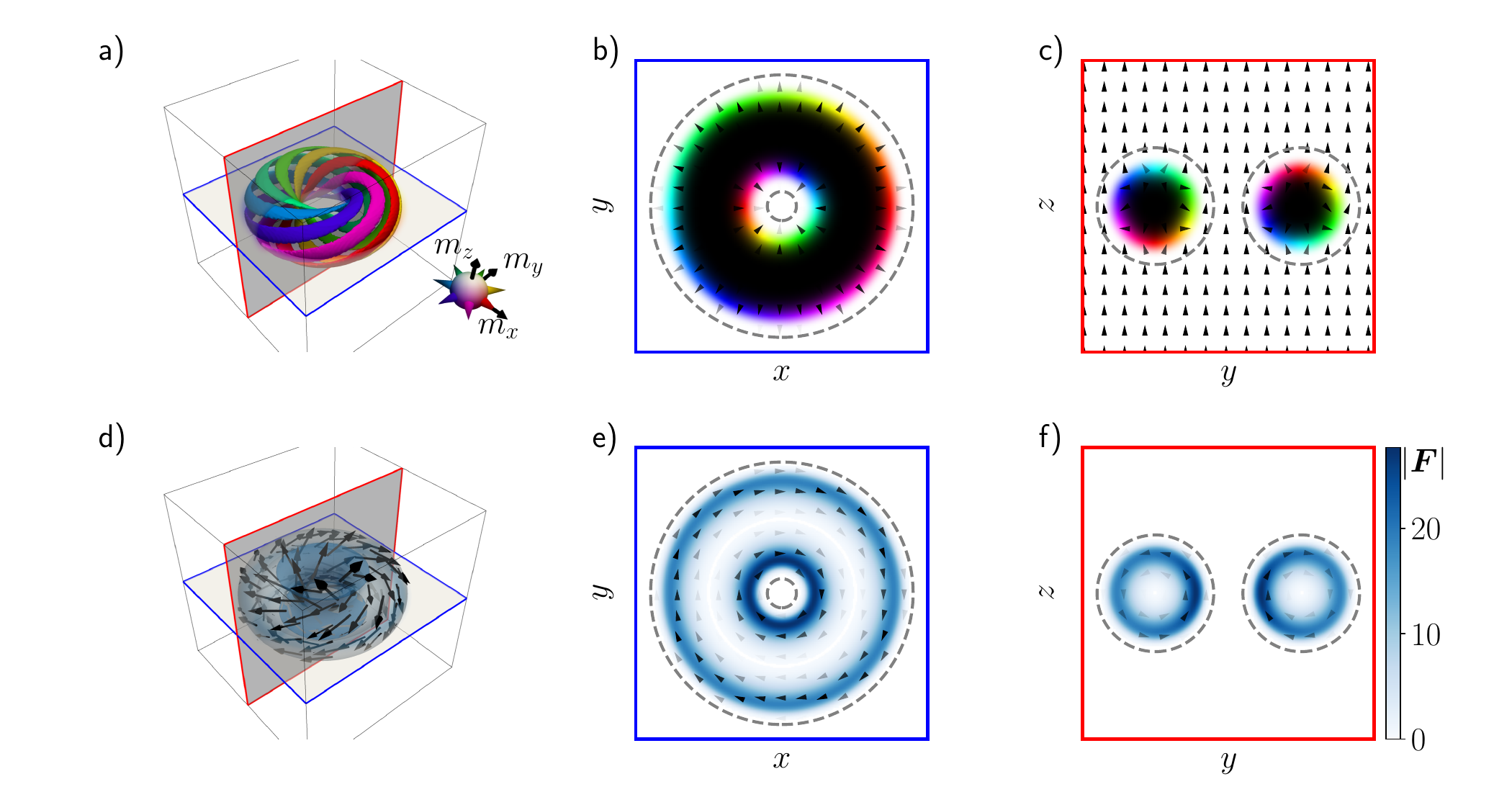}
	\caption{Sketch of a hopfion with compact support, where $\vect{m} = \hat{\vect{z}}$ outside of the toroidal region illustrated by the grey dashed circles. a)-c) Magnetisation configuration $\vect{m}$ and d)-e) corresponding emergent magnetic field $\vect {F}$. Panel a) shows a selection of preimages around the equator of $S^2$, and panel d) shows the emergent magnetic field, as well as various isosurfaces of constant $|\vect{F}|$. Panels b) and e) show the cross-sections in the plane of $z=0$, and panels c) and f) show cross-sections in the plane of $x=0$. The size of the hopfion relative to the boundaries is as in our benchmarking results.}
\label{fig:Hopfion}
\end{figure*}

In this work, we provide several implementations of the numerical calculation of the Hopf index based on the Whitehead equation in the finite-difference modelling using both a general Python script supplied in Ref.~\cite{SupplementaryPythonHopfIndex}, as well as an extension to the micromagnetic software \texttt{MuMax3}~\cite{Vansteenkiste2014} (available at \url{https://mumax.github.io}, the extensions are available in Ref.~\cite{SupplementaryMuMaxSource}), and we compare their accuracies.
We benchmark the different codes on a magnetic hopfion whose emergent magnetic field $\vect F$ is a vector-valued function with compact support, i.e.\ $\vect{F} \neq \vect{0}$ only in a finite region, see Fig.~\ref{fig:Hopfion}. For brevity, we will refer to it as a hopfion with compact support in the following.

\section{Challenges of Computing the Whitehead integral}

The reliance of Eq.~\eqref{eq:HopfIndex} on the gauge potential $\vect{A}$ introduces challenges. For a given vector field $\vect F$, the gauge potential $\vect{A}$ is not uniquely defined and varies under gauge transformations. 
Thus, the integrand itself, which one could refer to as ``Hopf density'', is also a gauge-dependent quantity. In particular, this means that, in contrast to the 2D skyrmion density described by Eq.~\eqref{eq:EmergentField}, which is gauge invariant, it is generally not meaningful to analyse and visualise the Hopf density, see App.~\ref{app:GaugeDependence}.

The Hopf index $H$, however, despite being computed using a gauge-dependent $\vect A$, must remain gauge-independent as a well-defined topological invariant. 
Here a very subtle point arises which can be easily overlooked in numerical computations, wherefore we discuss the seemingly smooth gauge invariance argument in detail.

A gauge transformation can be expressed as a shift of the gauge potential $\vect{A}$ by the gradient of a scalar function $\chi$, i.e.\
$ \vect{A}\rightarrow \vect{A'}= \vect{A} + \vect{\nabla} \chi$.
Note that the vector field $\vect{F}$ itself is gauge invariant, as the curl of a gradient vanishes.
Inserting the gauge transformed potential into Eq.~\eqref{eq:HopfIndex} yields
\begin{equation}
H' =  -\int_V \mathrm{d}^3 r \, \vect{F} \cdot \vect{A}' 
= H - \int_V \mathrm{d}^3 r \, \vect{F} \cdot \vect{\nabla}\chi.
\label{eq:gaugeH}
\end{equation}

To ensure gauge invariance, the last term in Eq.~\eqref{eq:gaugeH} must vanish. Using $\vect{F}\cdot \vect{\nabla}\chi = \vect{\nabla} \cdot (\chi \vect{F})$, as $\vect{\nabla} \cdot \vect{F} =0$, one can apply Gauss's theorem 
\begin{equation}
    \label{eq:chiF}
    \int_V \mathrm{d}^3 r \, \vect{F} \cdot \vect{\nabla}\chi
    = \oint_{\partial V} \mathrm{d}^2 r \,  \chi \, \vect{F} \cdot \vect{n},
\end{equation}
where $\partial V$ is the closed surface of the volume $V$ with the spatially varying surface normal $\vect n$.

Thus $H$ is only well-defined if for ``arbitrary" functions $\chi$ the surface integral in Eq.~\eqref{eq:chiF} vanishes.
To be more precise, for a given vector field $\vect F$, the integration volume, as well as $\chi$ and, consequently, $\vect A$, are not entirely arbitrary. For instance, in the derivation, we assumed that $\chi$ is differentiable.
Overall, for a well-defined Hopf index, the integration volume $V$ must be chosen such that either $\vect{F}\cdot \vect{n}=0$ on the boundary of the volume $\partial V$ or the function space for $\vect A$ must be appropriately restricted by the specific problem under consideration 
such that the surface integral in Eq.~\eqref{eq:chiF} vanishes for any gauge choice.

In many cases, the integration volume considered is the infinite space $\mathbb{R}^3$. However, numerical calculations of $H$ using Eq.~\eqref{eq:HopfIndex} typically require integration over finite, bounded volumes. Eq.~\eqref{eq:chiF} emphasises that comparison and benchmarking of the results of different numerical implementations, which may use different gauge choices, is only fair if the vector field $\vect{F}$ has compact support.
Then the simply-connected integration volume can be chosen such that outside $V$, and on its boundary, $\vect F\equiv \vect{0}$.~\footnote{
This is a stricter condition than that the far field of $\vect F$ converges 
to zero ``quickly enough''.}

As a side note we want to mention two cases going beyond the case of a simply-connected, finite-sized volume discussed above: In the case of a multiply-connected volume, such as a nanoring, a term that accounts for the holes in the volume must be added to Eq.~\eqref{eq:HopfIndex}, as discussed in Ref.~\cite{MacTaggart2019}.
For magnetic textures which are periodic along a certain axis, the Hopf index is computed per length, and the base space is $S^2 \times T^1$~\cite{Auckly2005,Jaykka2009,Kobayashi2013}. The Hopf index is then a $\mathbb{Z}^2$ invariant~\cite{Knapman2024}.

\section{Numerical implementations of the Hopf Index}
The calculation of the Hopf index requires the calculation of the emergent magnetic field $\vect{F}$, see Sec.~\ref{sec:F}. For this, we provide two different methods. In the first method, we directly compute the derivatives in Eq.~\eqref{eq:EmergentField} using finite-difference discretisation. In the second, the emergent field is interpreted geometrically as the solid angle subtended by magnetisation vectors over the lattice.

Having calculated $\vect{F}$, we discuss two different approaches by which the Hopf index $H$ can be calculated. In Sec.~\ref{sec:FiniteDifference}, we calculate the vector potential $\vect{A}$ explicitly and integrate $\vect{F} \cdot \vect {A}$ over space as in Eq.~\eqref{eq:HopfIndex}. In Sec.~\ref{sec:Fourier}, we circumvent the need to explicitly calculate $\vect{A}$ by transforming $\vect{F}$ to Fourier space~\cite{Liu2018}.

\subsection{Numerical Calculation of the Emergent Magnetic Field}
\label{sec:F}
The discretised emergent magnetic field can either be calculated using a finite-difference discretisation of Eq.~\eqref{eq:EmergentField} or using a solid angle representation, discussed below.
\subsubsection{Finite-Difference Method}
To calculate the emergent magnetic field using Eq.~\eqref{eq:EmergentField}, the derivatives need to be discretised to match the lattice discretisation.
In the provided implementations, derivatives of the magnetisation that appear in the vector field $\vect F$ are approximated and implemented using either the second-order accurate two-point central differences stencil discretisation\footnote{Please note that the convention in the literature is to call a numerical approximation of a derivative taking just the two nearest neighbours of a point into account a two-point stencil while including also the second nearest neighbours is called a five-point stencil.}, e.g.\
\begin{multline}
    \left. \frac{\partial \vect{m}(x,y,z)}{\partial y}\right|_{y=y_i} = \frac{\vect{m}(x,y_{i+1},z) - \vect{m}(x,y_{i-1},z)}{2 \Delta y} \\
    + \mathcal{O}[(\Delta y)^2],
\end{multline}
or a fourth-order accurate five-point stencil discretisation, e.g.\
\begin{equation}
\begin{aligned}
   \left. \frac{\partial \vect{m} (x,y,z)}{\partial y} \right|_{y=y_i} &= \frac{1}{12\Delta y} \bigl[ -\vect{m}(x,y_{i+2},z)\\
&+ 8\vect{m}(x,y_{i+1},z) - 8\vect{m}(x,y_{i-1},z)\\
   & + \vect{m}(x,y_{i-2},z)\bigr]+ \mathcal{O}[(\Delta y)^4],
\end{aligned}
\end{equation}
where $\Delta y$ is the side length of a finite-difference cell in the $y$-direction. The finite-difference method with two-point and five-point stencils gives errors in the calculated emergent magnetic field proportional to $\Delta^2$ and $\Delta^4$ respectively.

The function to obtain the emergent magnetic field has been implemented as \texttt{ext\_emergentmagneticfield\_twopointstencil} and \texttt{ext\_emergentmagneticfield\_fivepointstencil} for the two- and five-point stencils respectively. The corresponding functions to calculate the Hopf index are implemented as \texttt{ext\_hopfindex\_twopointstencil} and \texttt{ext\_hopfindex\_fivepointstencil}.

\subsubsection{Solid Angle Method}

As the out-of-plane component of the effective field is proportional to the in-plane skyrmion number density, we can calculate $\vect{F}$ using a method for calculating the skyrmion number developed by Berg and Lüscher~\cite{Berg1981}, which has been implemented in \texttt{MuMax3}~\cite{Kim2020}. In this method, the solid angle subtended by the magnetisation vectors is calculated.

\begin{figure}
    \centering
    \includegraphics[width=\linewidth]{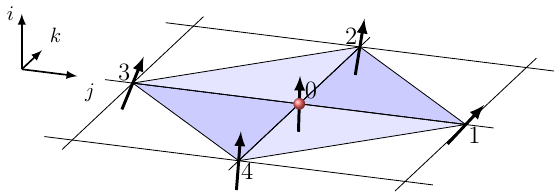}
	\caption{Illustration of the lattice sites contributing to the solid angles summed over to calculate the $i$-th component of the emergent magnetic field $\vect{F}$ at the lattice site $0$ (indicated by the red ball).}
    \label{fig:SolidAngleMethod}
\end{figure}

Here, for example, the $i$-th component of the emergent magnetic field at lattice site ``0'' is given by
\begin{equation}
    F_i= \frac{1}{16} (q_{012} + q_{023} + q_{034} + q_{041}),
    \label{eq:Fsolidangle}
\end{equation}
where $q_{lmn}$ is the solid angle subtended by the magnetisation vectors at lattice sites $l$, $m$, $n$ spanning the plane perpendicular to direction $i$, see Fig.~\ref{fig:SolidAngleMethod}. It is defined by
\begin{equation}
   q_{lmn} =  2 \arctan \left( \frac{\vect{m}_l \cdot (\vect{m}_m \times \vect{m}_n)}{1 + \vect{m}_l \cdot \vect{m}_m + \vect{m}_l \cdot \vect{m}_n + \vect{m}_m \cdot \vect{m}_n} \right).
\end{equation}
We have implemented this calculation of the emergent magnetic field in \texttt{MuMax3} as \texttt{ext\_emergentmagneticfield\_solidangle}.

\subsection{Direct Numerical Implementation of the Whitehead Equation}
\label{sec:FiniteDifference}
For the direct numerical implementation of Eq.~\eqref{eq:HopfIndex} in a micromagnetic solver which employs a finite difference discretisation of space such as \texttt{MuMax3}, a gauge choice must first be made, then all quantities need to be discretised to match the space discretisation.

We choose
\begin{equation}
\begin{aligned}
	A_x(x, y, z) &= \int_{-L_{y/2}}^y \mathrm{d}y' F_z(x, y', z), \\
	A_y(x, y, z) &= 0, \\
	A_z(x, y, z) &= -\int_{-L_{y/2}}^y \mathrm{d}y' F_x(x, y', z),
\end{aligned}
\label{eq:VectorPotential}
\end{equation}
where we integrate the vector potential along the $y$ direction in a box with volume $V=L_x L_y L_z$, having side lengths $L_x$, $L_y$ and $L_z$, see Fig.~\ref{fig:CoordinateSystem}.

We discretise Eq.~\eqref{eq:VectorPotential} as
\begin{equation}
    \label{eq:VectorPotentialDiscretised}
    A_x(x, y, z) \approx A_{x, i j k} = \sum_{j'=0}^{j-1} F_{z, ij'k} \, \Delta y,
\end{equation}
and analogously for $A_z$. Here $i$, $j$, $k$ are the indices of the cells in the $x$-, $y$-, and $z$-directions, respectively. For the points at which $j = 0$, we set the vector potential to zero. 
\footnote{Such a gauge also allows the computation of the Hopf index of magnetic structures which are periodic along $x$- or $z$-directions (but not along $y$).} The Hopf index is then approximated by
\begin{equation}
    H \approx -\Delta x \, \Delta y \, \Delta z \, \sum_{i, j, k} \vect{F}_{ijk} \cdot \vect{A}_{ijk}.
\end{equation}

\subsection{Implementation of the Whitehead Equation in Fourier Space}
\label{sec:Fourier}

In the Coulomb gauge $\vect{\nabla} \cdot \vect{A}=0$, the Whitehead equation in Fourier space can be expressed as~\cite{Liu2018}
\begin{equation}
    H = -\frac{i}{2\pi N_x N_y N_z} \sum_{\vect{k}} \frac{\vect{F}(-\vect{k}) \cdot [\vect{k} \times \vect{F}(\vect{k})]}{k^2},
\end{equation}
which eliminates the need to explicitly calculate the gauge potential $\vect{A}$. Here we have used the discretised Fourier transform, where the discretised Fourier transform of the emergent magnetic field $\vect F$ is given by
\begin{equation}
        \vect{F}(\vect{r}_{lmn}) = \sum_{\lambda \mu \nu} \vect{F}(\vect{k}_{\lambda \mu \nu}) e^{2\pi i \left( \frac{l\lambda}{N_x} + \frac{m\mu}{N_y} + \frac{n\nu}{N_z} \right) },
\end{equation}
and analogously for $\vect{A}$. $l$, $m$, and $n$ are integer indices labelling the lattice in real space and $\lambda$, $\mu$, and $\nu$ are integer indices labelling the lattice in Fourier space. $N_x$ is the total number of cells in the $x$-direction, and analogous for $y$ and $z$. We implement the calculation of the Hopf index using this method in \texttt{MuMax3} as \texttt{ext\_hopfindex\_latticefourier}, which first uses the solid angle method to calculate the real-space emergent magnetic field.

\subsection{Hopf index calculation in Finite-Element Micromagnetics}

To calculate the Hopf index Eq.~\eqref{eq:HopfIndex} using finite element methods (FEM), we first determine the vector field $\vect{F}$ by computing the spatial derivatives of the magnetisation's Cartesian components as in Sec.~\ref{sec:F}. 
In the Coulomb gauge, $\vect{\nabla}\cdot\vect{A}=0$,
the vector potential $\vect{A}(\vect{r})$ is then obtained by solving the Poisson equation 
$ \Delta \vect{A} = -\nabla \times \vect{F}$, using a hybrid finite-element/boundary-element algorithm similar to the methods described in Refs.~\cite{hertel_large-scale_2019, hertel_hybrid_2014}. 
We use linear finite elements, which entail an accumulation of discretisation errors due to multiple numerical differentiations in the right-hand side terms of Eq.~\eqref{eq:EmergentField} and the Poisson equation.

The software to determine the Hopf index in FEM micromagnetics was developed as an add-on to the open-source simulation software \texttt{tetmag}~\cite{hertel_tetmag_2023}. More details of this numerical method will be described in a separate work.

\section{Benchmarking Numerical Results for a Hopfion with Compact Support}

\begin{figure}
    \centering
    \includegraphics[width=0.8\linewidth]{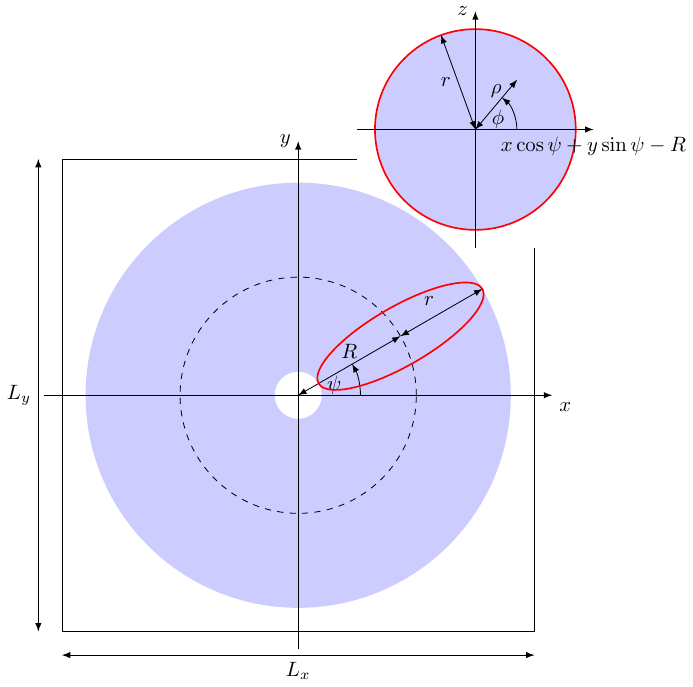}
    \caption{Illustration of the coordinate system used for the hopfion ansatz with compact support, where $\vect{m} = \hat{\vect{z}}$ outside of the torus with major radius $R$ and minor radius $r$. In our benchmarking, $R = 0.25$, $r = 0.2$, and $L_x = L_y = 1$.}
    \label{fig:CoordinateSystem}
\end{figure}

To benchmark and compare the different codes we use the following rotationally symmetric ansatz for hopfion with Hopf index $H = 1$, with major radius $R$ and minor radius $r$, see Fig.~\ref{fig:CoordinateSystem}.
The magnetisation vector field expressed in Cartesian coordinates is
\begin{equation}
    \vect{m}(x,y,z) = \begin{pmatrix}\cos\Phi \sin\Theta \\ \sin\Phi \sin\Theta \\ \cos\Theta \end{pmatrix},
\end{equation}
where the azimuthal and polar angles are given by
\begin{subequations}
\label{eq:Hansatz}
\begin{align}
    \Phi &= \psi - \phi \\
    \Theta & = 
            \begin{cases} 
            \pi \exp \left( 1 - \frac{1}{1 - (\rho/r)^2} \right), 
            & \text{ if } \rho < r \\ 0, 
            & \text { if } \rho \geq r.
            \end{cases}
\end{align}
\end{subequations}
The auxiliary radial coordinate $\rho$, and auxiliary angular parameters $\phi$ and $\psi$ that sweep around the minor radius $r$ and major radius $R$ are respectively given by
\begin{subequations}
\begin{align}
     \rho &= \sqrt{z^2 + (x\cos\psi + y\sin\psi - R)^2},\\
	\phi &= \arctan \left( \frac{z}{x\cos\psi + y\sin\psi - R} \right), \\
   \psi &= \arctan(y / x).
\end{align}
\end{subequations}
The magnetisation texture of this ansatz is illustrated in Fig.~\ref{fig:Hopfion}.

For this ansatz, the magnetisation outside a toroidal region with the major radius $R$ and the minor radius $r$ points strictly along the $z$ direction, and thus the emergent magnetic field $\vect F$ is zero.
To ensure gauge-independent results of our Hopf index calculation, we choose the integration volume such that it encloses the entire torus, i.e.\  $\vect F \equiv 0 $ on the integration boundary $\partial V$. Another advantage of the ansatz of Eq.~\eqref{eq:Hansatz} is that the magnetisation is infinitely differentiable, i.e.\ it is smooth.

In App.~\ref{app:movinghopfion} we also discuss an example of a 3D moving hopfion, which also shows the effect of the gauge dependence of the Hopf index when only a finite integration region is considered.

\subsection{Comparison of the Accuracy of the Hopf Index Calculations}

\begin{figure}
    \centering
    \includegraphics[width=\linewidth]{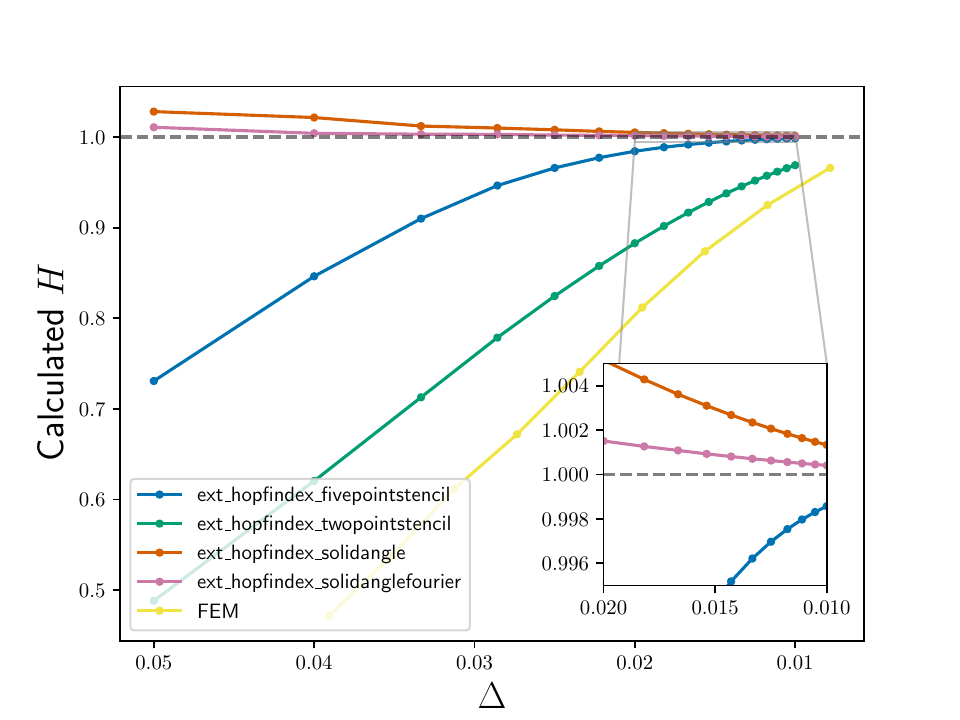}
	\caption{Calculated Hopf index of the $H = 1$ hopfion ansatz, Eq.~\eqref{eq:Hansatz}, with major radius $R = 0.25$ and minor radius $r = 0.2$ in a cubic system as a function of cell discretisation side length $\Delta$ for different numerical methods.
 }
    \label{fig:MethodsComparison}
\end{figure}

To assess the accuracy of the various methods to calculate the Hopf index, we plot it as a function of the side length of the discretisation cells $\Delta$ for the cubic system in Fig.~\ref{fig:MethodsComparison}. 
For the finite element results also a cubic mesh was used.
 We benchmark our codes on the Hopfion ansatz, Eq.~\eqref{eq:Hansatz}, with major radius $R = 0.25$ and minor radius $r = 0.2$.
The cubic integration box has a side length of $L = N \Delta = 1$  where $N$ is the number of cells in each direction. While we work in dimensionless units here, $L$ is arbitrarily rescalable. E.g.\ with $L=\SI{100}{\nano\metre}$, $ R=\SI{25}{\nano\metre}$, $ r=\SI{25}{\nano\metre}$ and $\Delta = \SI{2}{\nano\metre}$ would imply $N= 50$ and the accuracy ranges from $\sim 80\%$ to $\sim 99.8\%$ for the different methods.

We find that all methods converge towards the ideal value of $H = 1$ as the discretisation becomes finer. The Hopf index calculation for the case that $\vect{F}$ is calculated using the solid angle method is more accurate and converges faster than both the finite-difference derivative method and the finite element method. This enhanced accuracy is 
due to the solid angle method's efficiency in covering the unit sphere with fewer lattice points, even at coarser discretisation.
In contrast, derivative-based approaches require finer resolution to achieve comparable accuracy~\cite{Berg1981,Kim2020}.

\subsection{Computational Speed of the Hopf Index Calculation}

\begin{figure}
    \centering
    \includegraphics[width=\linewidth]{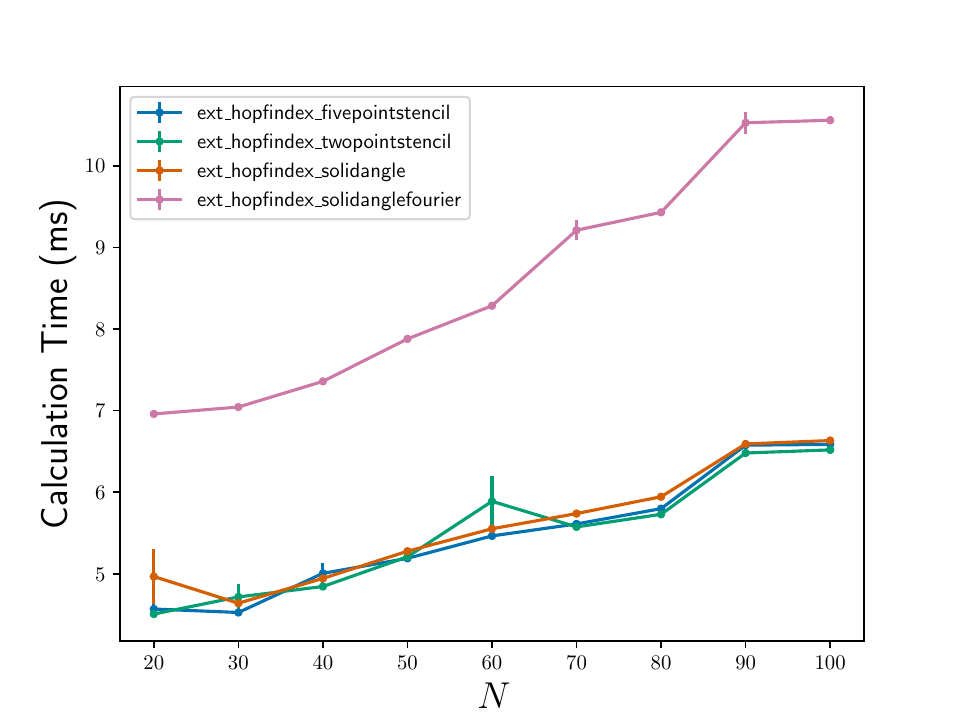}
	\caption{Time required to calculate the Hopf index as a function of the side length of the cubic system for the various methods used to calculate the Hopf index, averaged over 100 calculations. The shaded regions show the standard error in the computational time.}
    \label{fig:Benchmarking}
\end{figure}

The methods to calculate the Hopf index are GPU-accelerated, and the presented results took on the order of milliseconds to execute on a modern GPU. In Fig~\ref{fig:Benchmarking}, we show the time required to calculate the Hopf index for our compact hopfion ansatz as a function of the number of cells $N$ of the discretisation grid. The calculation was run on an Nvidia GeForce RTX 3070.

We see that all methods take approximately the same amount of time to compute the Hopf index, except for the Fourier space method discussed in Sec.~\ref{sec:Fourier}, which takes longer. Although the Fourier method does not require the explicit calculation of the vector potential $\vect{A}$, we find that calculating the Fourier transform of $\vect{F}$ is more computationally expensive. Because of this, the real-space lattice method is likely to be more useful in most applications, specifically \texttt{ext\_hopfindex\_solidangle}. The method using the calculation of the field in Fourier space is, however, using the Coulomb gauge, and thus allows for cross-checking the calculated Hopf index in different gauges.

\section{Discussion and Conclusion}

In this work, we have provided and benchmarked multiple implementations of the numerical calculation of the Whitehead formula Eq.~\ref{eq:HopfIndex} to compute the Hopf index of a magnetic texture. We provide both \texttt{MuMax3} extensions and general Python codes to calculate the Hopf index.
Among the methods provided, we find that typically the solid angle method implemented as the \texttt{MuMax3} extension \texttt{ext\_hopfindex\_solidangle} to calculate the emergent magnetic field $\vect F$ and thus the Hopf index $H$ is more accurate than the finite-difference method, and should be preferred.

We pointed out common pitfalls and emphasised that the vector potential $\vect A$ and the \enquote{Hopf density} $\vect F \cdot \vect A$ is gauge dependent, and thus the integration region must be carefully considered when calculating the Hopf index of a magnetic texture.
In a finite system, an ideal numerical value of $1$ for the Hopf index can only be obtained for a magnetic texture with compact support.

To conclude, our work eases the topological analysis of 3D textures while maintaining high accuracy and computational efficiency.

\section{Data and Code Availability}
The finite difference scripts used to generate the results in this study are available at \url{https://zenodo.org/records/14003272}~\cite{SupplementaryScripts}. 
The \texttt{MuMax3} codes to calculate the Hopf index are planned to be added as extensions in a future release of \texttt{MuMax3} and are available at \url{https://zenodo.org/records/14006428}~\cite{SupplementaryMuMaxSource}. An additional Python implementation of the \texttt{MuMax3} extensions is available at \url{https://zenodo.org/records/14007386}~\cite{SupplementaryPythonHopfIndex}.

\section{Acknowledgements}
We thank Jan Masell, Markus Garst, Nikolai Kiselev,  and Volodymyr Kravchuk for fruitful discussions.
We acknowledge funding from the German Research Foundation (DFG) Project No.~320163632 (Emmy Noether),
Project No.~403233384 (SPP2137 Skyrmionics), Project No.~278162697 (SFB 1242, project B10), Project No.~405553726 (CRC/TRR 270, project B12)
and Project No.~505561633 in the TOROID project co-funded by the French National Research Agency (ANR) under Contract No. ANR-22-CE92-0032.
J.~L.\ is supported by the Research Foundation – Flanders (FWO) through senior postdoctoral research fellowship No.~12W7622N.

\appendix

\section{Gauge Dependence of the Calculated Hopf Index}
\label{app:GaugeDependence}
\begin{figure}[tb]
    \centering
    \includegraphics[width=\linewidth]{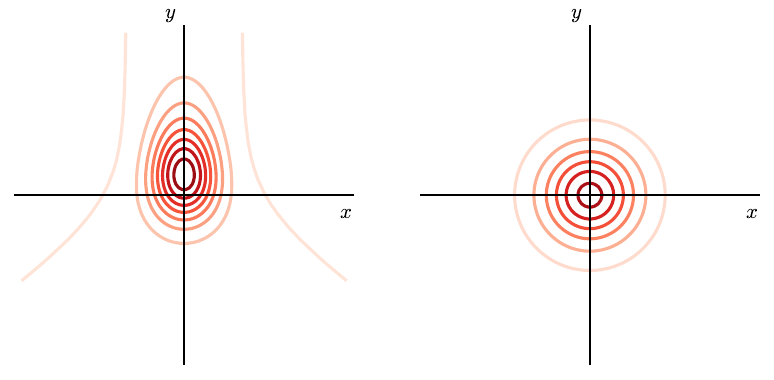}
    \caption{Contours of constant Hopf density $\vect{A} \cdot \vect{F}$ of a hopfion texture constructed by stereographic projection in the $z = 0$ plane for the gauge choice of Eq.~\ref{eq:VectorPotential} (left) and the Coulomb gauge used in the Fourier space method (right).}
    \label{fig:HopfDensity}
\end{figure}

In the main text, we have emphasised the gauge dependence of the ``Hopf density'' $\vect F \cdot \vect A$. 
In Fig.~\ref{fig:HopfDensity}, we show contours of constant Hopf density for two different gauges - the gauge of Eq.~\eqref{eq:gaugeH} and the Coulomb gauge. The spatial distribution of the contours illustrates the gauge dependence of the calculated value of the Hopf index when the integration region is not carefully considered. 

Here, the hopfion ansatz used is the standard stereographic projection (see e.g.~Ref.~\onlinecite{Liu2020}) to allow for an analytical expression of the Hopf density~\cite{Guslienko2023}. However, this Hopfion ansatz does not have compact support, i.e.,~$\vect F$ only converges to zero at infinity increasing the challenges of choosing integration regions for the calculation of the Hopf index.

For completeness, the Hopf density expression (\texttt{ext\_hopfindexdensity\_*}) as well as the possibility to export the emergent magnetic field (\texttt{ext\_emergentmagneticfield\_*}) are also included in the implementations.

\section{Moving Hopfions}
\label{app:movinghopfion}

It has been shown that vortex rings in a system with symmetric exchange and uniaxial anisotropy can propagate along the easy axis~\cite{Papanicolaou1993,Cooper1999,Sutcliffe2007}. 
Using our implementations, we calculate the Hopf index for a magnetic vortex ring (which does not have a compact support) in a finite region for such a system to analyse whether or not the propagating texture is topologically trivial.

The results are shown in Fig.~\ref{fig:Propagation}, as well as in the Supplementary Movie. It is important to note that, when the hopfion structure and its Hopf density are almost entirely within the integration region (marked as a green rectangle), the Hopf index is approximately the ideal value of $1$ for all methods except for that using the less accurate two-point stencil.

At the times for which the hopfion and/or its Hopf density is crossing into and out of the highlighted region, however, there is a significant discrepancy between the methods. The key difference, despite numerical accuracy, originates from the gauge choice, in agreement with Fig.~\ref{fig:HopfDensity}.
This further highlights the need to choose the integration region carefully such that the hopfion and its density are (almost) entirely localised within the region of interest.

\begin{figure}[tb]
    \centering
    \includegraphics[width=\linewidth]{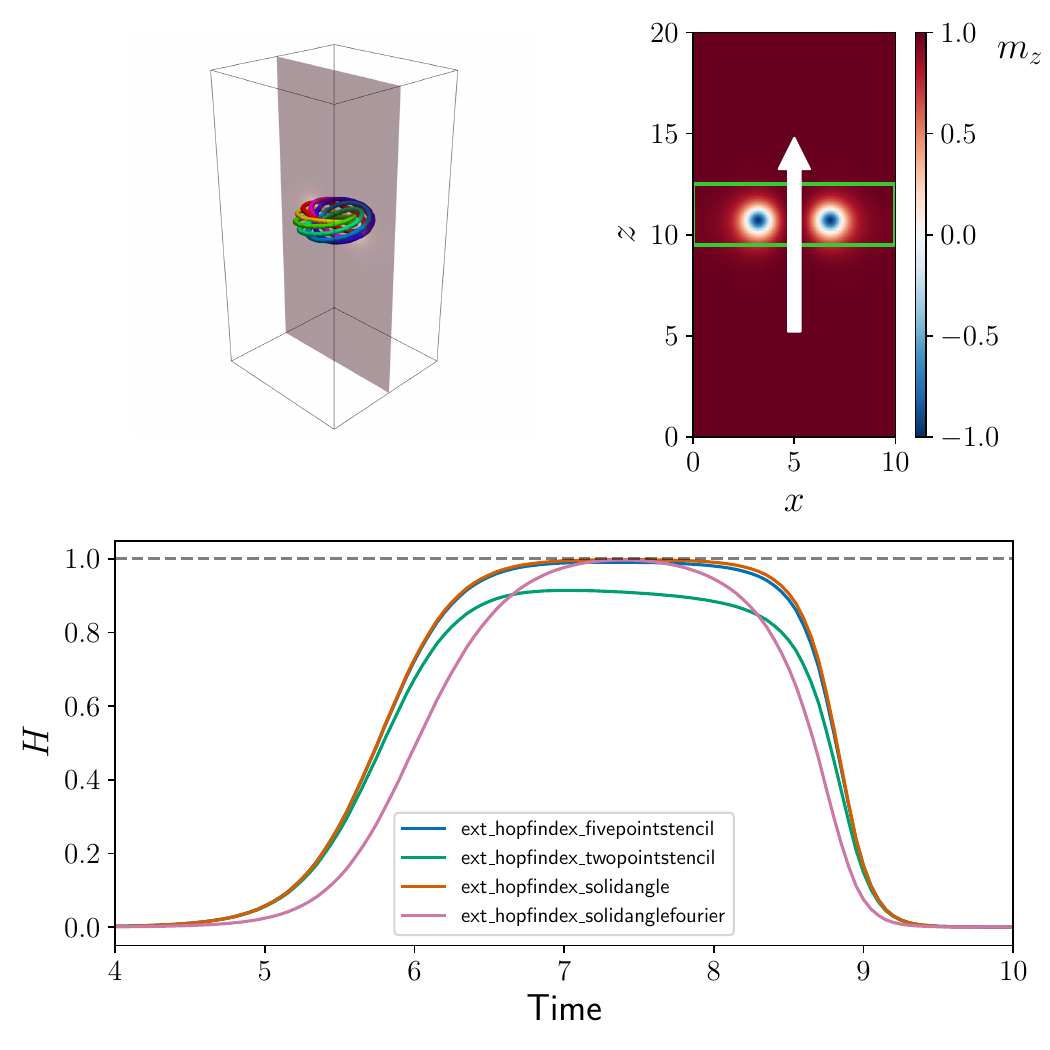}
	\caption{Calculation of the Hopf index of a propagating hopfion in a subregion of the system. Top left: Preimages of the 3D magnetisation texture, with a slice in the $xz$-plane. Top right: $z$-component of the magnetisation of the slice shown in the top-left image. The green rectangle illustrates the region in which the Hopf index is calculated, and the white arrow indicates the direction of propagation of the hopfion. 
 Bottom: Hopf index over time in the subregion calculated using the various methods discussed in the main text.}
    \label{fig:Propagation}
\end{figure}

\providecommand{\noopsort}[1]{}\providecommand{\singleletter}[1]{#1}%

\end{document}